\documentclass[aps,prx,onecolumn,english,superscriptaddress]{revtex4-2}

\usepackage{natbib}
\usepackage{amsfonts, amsmath, amssymb, mathrsfs, mathtools, amsthm, bbm, bm}
\usepackage{braket}
\usepackage{float}
\usepackage{afterpage}
\usepackage{color, soul}
\usepackage[version=4]{mhchem}
\usepackage{graphicx}
\usepackage[
    labelfont=bf, 
    font=small,
    justification=justified,
    format=plain
    ]{caption}
\usepackage[font=footnotesize,justification=justified]{subcaption}
\usepackage{comment}
\usepackage{tikz}
\usetikzlibrary{shadows,arrows.meta,positioning,backgrounds,fit,shapes}
\usepackage{quantikz}
\usepackage{booktabs, tabularx, colortbl}
\usepackage{quantikz}
\usepackage{blochsphere}
\usepackage{epstopdf}
\usepackage{adjustbox}
\usepackage[linesnumbered,ruled,vlined]{algorithm2e}

\AppendGraphicsExtensions{.tga}

\makeatletter
\def\BState{\State\hskip-\ALG@thistlm}
\makeatother

\newcolumntype{Y}{>{\raggedleft\arraybackslash}X}

\theoremstyle{definition}

\makeatletter
\renewcommand*\env@matrix[1][*\c@MaxMatrixCols c]{%
  \hskip -\arraycolsep
  \let\@ifnextchar\new@ifnextchar
  \array{#1}}
\makeatother

\usepackage[bookmarks=false]{hyperref}

\begin{document}

\title{Practical Log-Depth Quantum State Preparation and Circuit Verification via Tree Tensor Network Compilation}

\author{Angus Mingare}
\email{angus.mingare.22@ucl.ac.uk}
\affiliation{Centre for Computational Science, Department of Chemistry, University College London, WC1H 0AJ, United Kingdom}

\author{Peter V. Coveney}
\affiliation{Centre for Computational Science, Department of Chemistry, University College London, WC1H 0AJ, United Kingdom}
\affiliation{Advanced Research Computing Centre, University College London, WC1H 0AJ, United Kingdom}
\affiliation{Informatics Institute, University of Amsterdam, Amsterdam, 1098 XH, Netherlands}

\date{May 5, 2026}

\begin{abstract}
Matrix product states provide efficient classical descriptions of quantum systems that may be useful as reference states for quantum algorithms such as quantum phase estimation and quantum-selected configuration interaction. Shallow circuit constructions for loading matrix product states onto quantum computers is necessary for this to be practical on near-term hardware. We present a decomposition of matrix product states to log-depth quantum circuits via a simple tree tensor network renormalisation procedure. Our method exposes an explicit parameter which can be used to trade a small amount of fidelity for large savings in circuit depth. We extend this decomposition to the case of matrix product operators allowing us to construct log-depth and ancilla-free circuits to calculate overlaps of the form $\left |\braket{\phi|U|\psi}\right |^2$. In particular, we demonstrate an interpretation of these circuits as \emph{verifier circuits} with application to circuit-level device calibration.
\end{abstract}

\maketitle

\section{Introduction}

Efficient reference state preparation remains a key challenge in quantum algorithms, particularly those with applications to chemistry such as quantum phase estimation \cite{kitaev1995quantum} and quantum-selected configuration interaction \cite{kanno2023quantum, robledo2025chemistry}, where the quality of the initial state directly impacts performance \cite{fomichev2024initial, sennane2026robustness, nakagawa2024adapt}. Tensor networks, specifically matrix product states (MPS), provide efficient classical representations of quantum states \cite{orus2014practical}. Methods such as the density-matrix renormalisation group (DMRG) \cite{white1993density, schollwock2011density} yield high-quality approximate solutions to the groundstate problem in MPS form, which are then ideal candidates as reference states for quantum algorithms \cite{rudolph2023synergistic, jaderberg2026variational}. However, loading these representations onto quantum computers is a non-trivial problem, especially for near-term quantum devices where circuit depth is severely limited by noise \cite{preskill2018quantum, bharti2022noisy}. Common MPS to circuit mappings for near-term devices use layers of staircase circuits \cite{schon2005sequential, ran2020encoding} which may be prohibitively deep when scaling up to tens to hundreds of qubits. 

In this work, we introduce a method for compiling MPS into quantum circuits with logarithmic depth by renormalising the state into a binary tree tensor network (TTN). This is achieved through a sequence of local merges and singular value decompositions, yielding a hierarchical structure that can be directly mapped to a circuit composed of multi-qubit gates. The resulting circuits exhibit depth $\mathcal{O}(\log{N})$ for $N$ qubits, at the cost of moderately increased gate locality scaling with the bond dimension.

We analyse the effect of truncation during the renormalisation procedure and show that the resulting infidelity grows at most linearly with system size, with a prefactor determined by the discarded singular values. Numerical results confirm that this prefactor remains small in practical settings, enabling accurate approximations even for large systems. This approximation significantly improves transpiled circuit depths and makes this method practical for near-term hardware. 

We further show that the same construction naturally extends to operator verification. By mapping a matrix product operator (MPO) to an MPS via vectorisation, our method yields a quantum circuit that evaluates overlaps of the form $\left | \braket{\phi|U|\psi}\right |^2$ as a single computational basis amplitude. This provides a structured alternative to standard techniques such as the swap test, enabling efficient circuit-level fidelity estimation and calibration when $U$ admits a low bond-dimension representation.

We provide necessary background and comment on related work in Section \ref{section:background}. The MPS to circuit mapping procedure for exact and approximate decompositions is given in Section \ref{section:mps}. Finally, Section \ref{section:vc} provides the extension to MPOs with a focus on verifier circuit constructions. 

\section{Background}
\label{section:background}

\subsection{Tensor Networks}

\begin{figure}[htbp]
    \centering
    \includegraphics[width=0.6\textwidth]{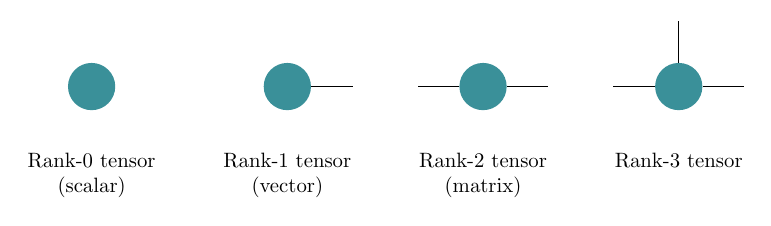}
    \caption{Basic tensors depicted in the tensor network graphical language. A tensor is represented by a node with a number of legs equal to the tensor rank.}
    \label{fig:basic_tn}
\end{figure}

A tensor network is a representation of a quantum system by a collection of low-dimension tensors \cite{orus2014practical}. Graphically, a tensor is represented by a node with a number of legs equal to the tensor's rank. Some examples are shown in Figure \ref{fig:basic_tn}.

\begin{figure}[htbp]
    \centering
    \includegraphics[width=0.35\textwidth]{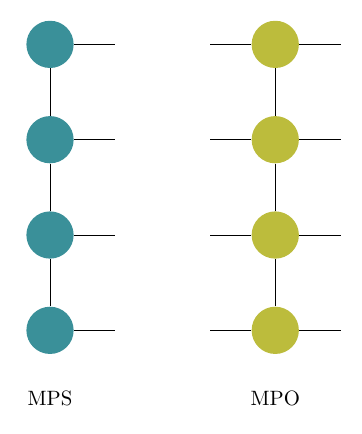}
    \caption{A matrix product state (MPS) and a matrix product operator (MPO), each with 4 sites.}
    \label{fig:mps_mpo}
\end{figure}

A tensor contraction, or summing over a shared index, is depicted by connecting the associated legs in the tensor network diagram. A matrix product state (MPS) is a particular tensor network structure used to represent quantum states. More specifically, we take a one dimensional chain of matrices with entries in $\mathbb{C}^d$. Similarly, a matrix product operator (MPO) is a one dimensional chain of matrices with operator entries $\mathbb{C}^{d\times d}$. An MPS and an MPO are shown in Figure \ref{fig:mps_mpo}. 

In this work, the external legs have dimension $d=2$ as we are considering a system of $N$ qubits. The maximum dimension of the internal legs is referred to as the bond dimension of the tensor network and is denoted $\chi$. The bond dimension, $\chi$, controls the time and memory complexity of tensor network algorithms. To represent exactly an arbitrary quantum system requires $\chi$ to grow exponentially with $N$. However, very good approximations are often achieved with a small, fixed bond dimension. 

The density matrix renormalisation group (DMRG) algorithm is a variational algorithm over the manifold of MPSs with bond dimension at most $\chi$ designed to find approximate ground state solutions given a system Hamiltonian represented as an MPO \cite{schollwock2011density}. A computationally inexpensive DMRG solution may provide a useful reference state for more accurate quantum algorithms such as quantum phase estimation or quantum-selected configuration interaction. 

\subsection{Quantum Circuit Synthesis}

\begin{figure}[htbp]
    \centering
    \includegraphics[width=0.6\textwidth]{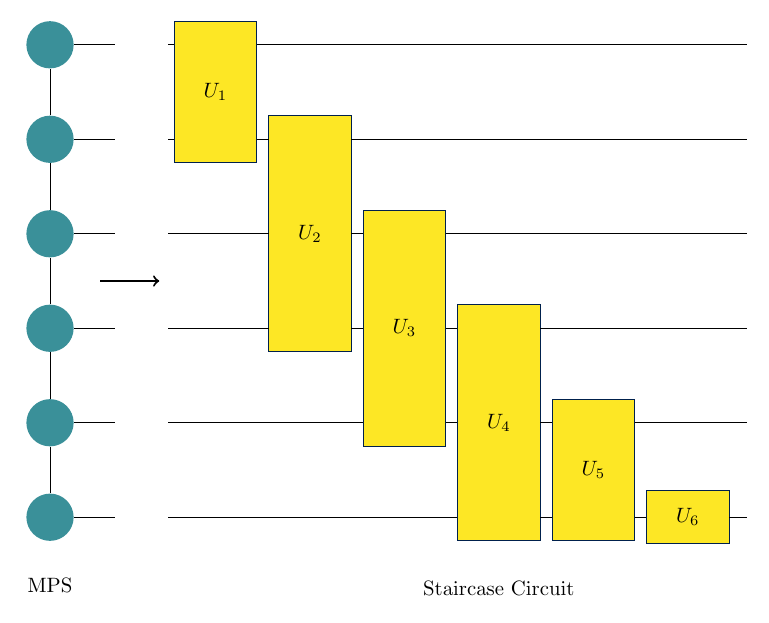}
    \caption{An MPS with $N=6$ and $\chi=4$ in canonical form is mapped directly to a linear-depth staircase circuit with maximum gate size $1+\log{(\chi)}$.}
    \label{fig:staircase}
\end{figure}

To use MPSs as reference states for quantum algorithms we need to be able to load the state onto a quantum computer. A simple way to do this that is suggested by the graphical representation, is to decompose the MPS into a staircase circuit as shown in Figure \ref{fig:staircase} \cite{schon2005sequential}. 

For this mapping to result in a valid quantum circuit, the MPS must first be put into canonical form with the orthogonality centre at one end. In this form, all tensors (with the exception of the orthogonality centre) are isometries from one internal bond to the physical bond and other internal bond. A tensor network can be placed into canonical form with an orthogonality centre at a chosen site through a sweep of singular value decompositions. Isometries can be embedded in unitary matrices with the correct action on the input state $\ket{0}$ by padding the input dimension with additional orthonormal vectors. Because the full MPS represents a normalised quantum system, the non-orthogonal site when interpreted as a vector can also be embedded in a unitary gate.

The staircase circuit decomposition requires gates acting on $1+\log\chi$ qubits and has total depth scaling as $\mathcal{O}(N)$ where potentially large pre-factors will come from the transpilation overhead associated with decomposing multi-qubit gates to particular architectures. Some approaches mitigate this transpilation overhead by replacing the single staircase circuit with an ansatz of layers of staircase circuits containing only two-qubit gates. The exact gates in the ansatz can be determined by analytical decompositions or numerical optimisation \cite{rudolph2023synergistic, rudolph2024decomposition}. A further saving can be obtained by beginning the staircases at the midpoint of the MPS rather than one end. This approximately adds a factor of $\frac12$ to the total circuit depth, however the asymptotic scaling remains $\mathcal{O}(N)$.

\begin{figure}[htbp]
    \centering
    \includegraphics[width=0.6\textwidth]{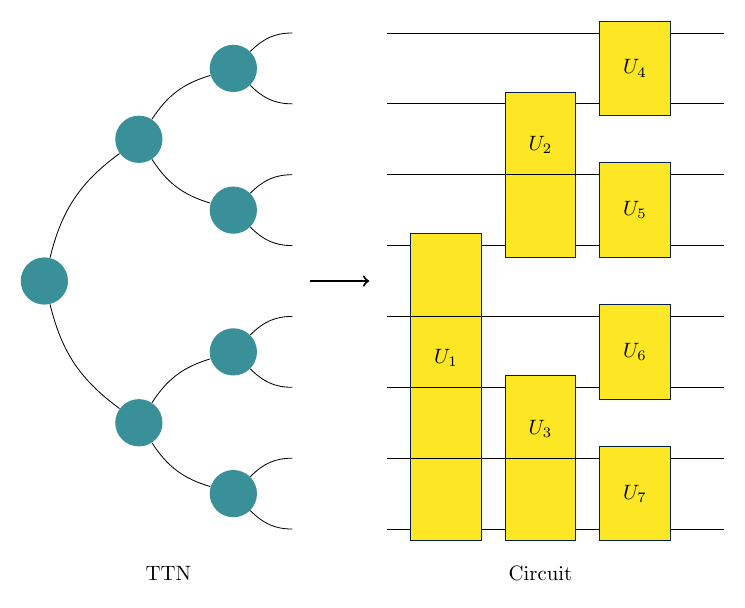}
    \caption{A 3-layer TTN with $\chi=2$ and orthogonality centre at the root is mapped directly to a log-depth staircase circuit.}
    \label{fig:logcirc}
\end{figure}

From the graphical representation, there is no obvious way to decompose an MPS to a quantum circuit with asymptotically better depth. However, there is another tensor network structure where this is possible. A tree tensor network (TTN), as shown in Figure \ref{fig:logcirc}, when appropriately canonicalised can be mapped directly to a log-depth quantum circuit using the same isometry to unitary embedding approach. 

TTNs are not as widely studied as the more ubiquitous MPSs and algorithms such as a TTN-DMRG are underdeveloped. Therefore, to exploit advantages of both structures, we must consider renormalisation strategies between MPSs and TTNs. We present the renormalisation procedure used in this work in Section \ref{section:procedure}.

\subsection{Related Work}

Our work is most closely related to Malz et al. \cite{malz2024preparation} and the follow-up by Scheer et al., \cite{scheer2025renormalization}, both of which present MPS to TTN renormalisation schemes for log-depth circuit preparation. A key theoretical result of Malz et al. is a lower bound showing that exact preparation of translation-invariant normal MPS requires depth $\mathcal{\omega}(\log N)$ in the local-gate model; their construction achieves $\mathcal{O}(\log (N/\epsilon))$ depth for error $\epsilon$, which is optimal in that setting. Our construction operates in the same spirit but differs in two important respects. First, we expose an explicit truncation parameter giving direct, tuneable control over the fidelity/depth trade-off, and we benchmark this concretely on realistic hardware topologies. Second, we extend the decomposition to matrix product operators, yielding log-depth ancilla-free verifier circuits for computing overlaps $|\braket{\phi|U|\psi}|^2$, with application to circuit-level device calibration, a direction not pursued in either of those works. 

On this second point, the concept of verifier circuits built from matrix product operators was introduced by Mingare et al. \cite{mingare2024quantum}. We extend this idea by demonstrating two generalisations. First, that these circuits can be constructed using any MPS to circuit mapping by first vectorising the MPO. Second, that the resulting circuits compute overlaps and thus have utility beyond circuit verification.

Other MPS to circuit construction methods achieve log-depth and even constant-depth circuits for preparing an MPS by exploiting measurements and classical feed-forward \cite{malz2024preparation, smith2024constant}. However, high-fidelity mid-circuit measurements and low-latency classical control remain challenging for current and near-term hardware. Furthermore, the non-unitarity of these approaches may make them inappropriate for integration into larger-scale quantum workflows. 

Variational methods for loading MPSs onto quantum computers are also being explored. The work by Robertson et al. \cite{robertson2025approximate} is a recent example of optimising shallow parametric quantum circuits to represent MPSs that has seen some success in practice \cite{jaderberg2026variational}. In contrast, our decomposition is analytical, requiring no classical optimisation. 

\section{Log-Depth State Preparation}
\label{section:mps}

\subsection{Renormalisation Procedure}
\label{section:procedure}

Our method decomposes an arbitrary MPS into a quantum circuit by first renormalising the tensor network structure into a binary tree. It proceeds as follows. We consider an MPS with $N$-sites and bond dimension $\chi$. If necessary, the tensors of the MPS are padded with zeroes so that $\chi = 2^k$ for some $k\in\mathbb{N}$ so that each bond can be interpreted in terms of qubits. At each iteration, we block together neighbouring sites in the MPS and perform a singular value decomposition (SVD) using the virtual bonds as input legs and the physical bonds as output legs. This creates an isometry while moving the non-orthogonal component back to the next layer. The number of singular values retained during the SVD is the parameter that will control the trade-off between circuit depth and fidelity. Similar to the bond dimension $\chi$ we ensure that we only ever truncate the SVD to a power of $2$ to retain the quantum circuit interpretation. 

\begin{figure}[htbp]
    \centering
    \includegraphics[width=\textwidth]{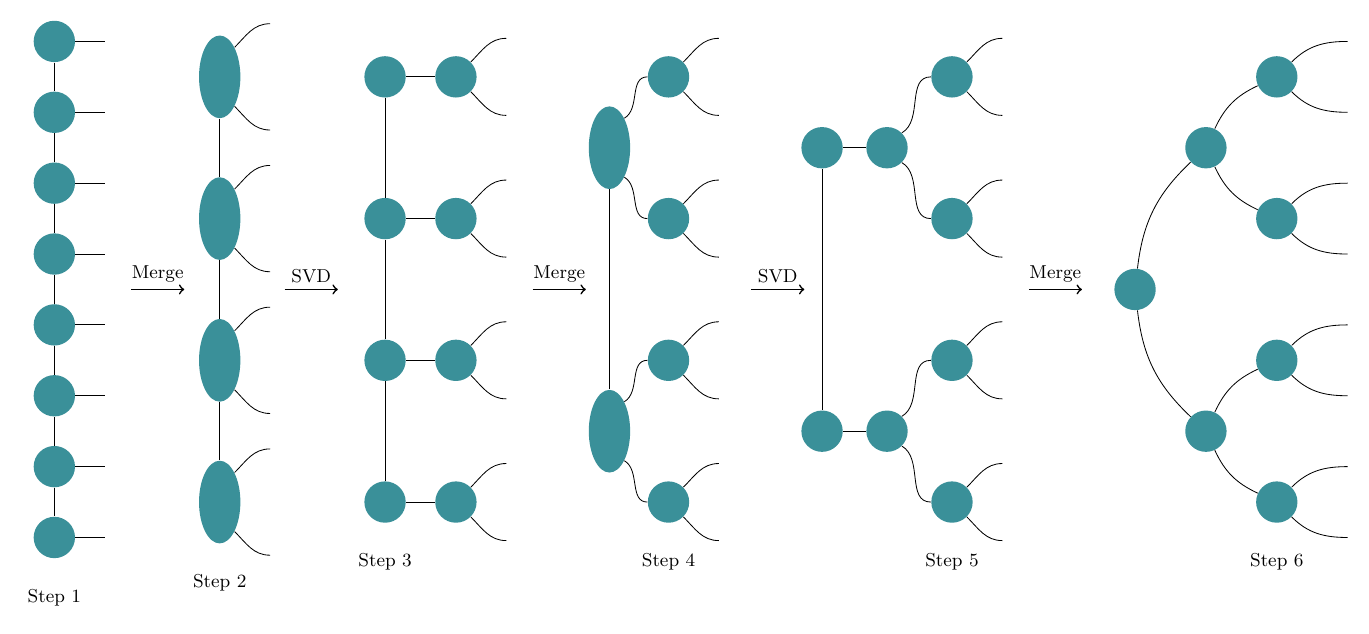}
    \caption{A series of local site merges and singular value decompositions brings an MPS into the form of a TTN.}
    \label{fig:mps_to_circ}
\end{figure}

The merge-SVD sequence is repeated to form $\lceil \log_2(N)\rceil$ layers with decreasing number of nodes. The final layer consists of a single non-orthogonal tensor while all other nodes are isometries moving from the root of the tree to the leaves. These isometries can then directly be embedded into unitary gates to form the quantum circuit decomposition. Because the MPS state is assumed to be a properly normalised quantum state, the final non-orthogonal node (which may be interpreted as a vector) can also be embedded into a unitary. The MPS to TTN renormalisation procedure is shown in Figure \ref{fig:mps_to_circ}. The final TTN to circuit interpretation is the same as that depicted previously in Figure \ref{fig:logcirc}.

\subsection{Complexity Analysis}
\label{section:complexity}

If the SVDs are not truncated, the bond dimension required to renormalise the MPS as a binary tree grows with each layer. For sufficiently large $N$, the maximum bond dimension will be $\chi^2$. Hence a single node in the tree can have output dimension at most $\chi^4$, corresponding to a gate acting on $4\log{(\chi)}$ qubits, independent of $N$. 

To determine when this bound is achieved we consider an SVD in layer $k$ of the binary tree. For a node not on the boundary of the tree, the SVD will be performed on a matrix with shape $(p_0p_1, \chi^2)$ where $p_0,p_1$ are the dimensions of the physical indices at that site. In layer $k$ a node will be connected to at most $2^k$ leaves of the tree and hence $p_0p_1 \leq 2^{2^k}$. Therefore, to achieve the maximum bond dimension $\chi^2$ we require,

\begin{equation}
    2^{2^k} \geq \chi^2,
\end{equation}

from which it follows

\begin{equation}
    k \geq 1 + \log{(\log{(\chi)})}.
\end{equation}

\begin{figure}[htbp]
    \centering
    \includegraphics[width=0.6\textwidth]{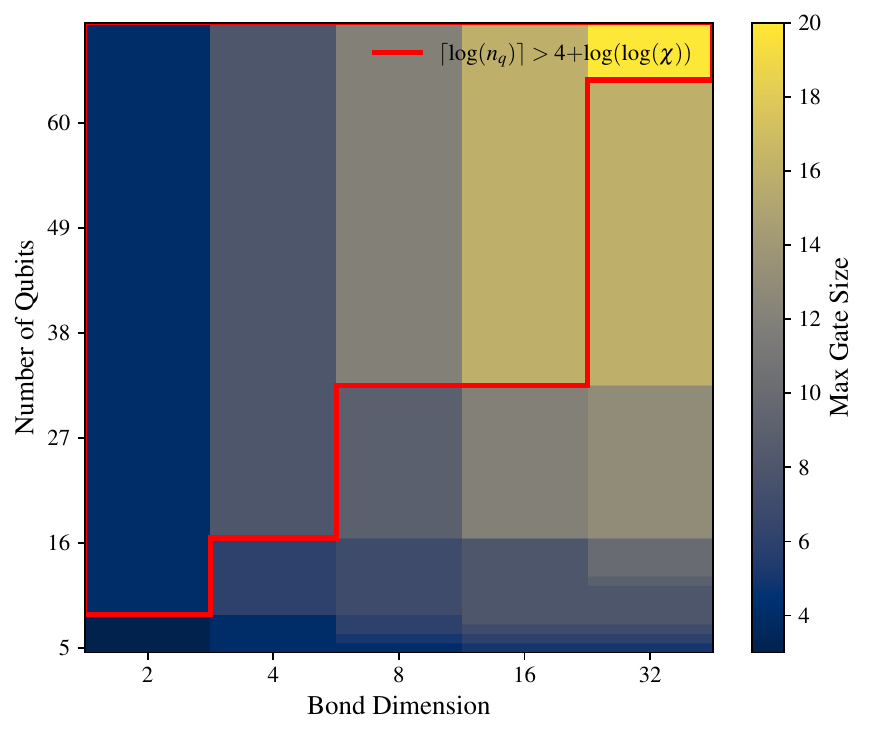}
    \caption{Maximum gate size plotted as a function of MPS bond dimension, $\chi$, and the number of sites. For a given $\chi$ there is a value of $N$ above which the maximum gate size saturates to $4\log{(\chi)}$, independent of $N$. This boundary is shown in red.}
    \label{fig:gate_size_heatmap}
\end{figure}

All that remains to establish the bound is to determine the layer closest to the root of the tree where it is possible for a node to have two output legs with dimension $\chi^2$. Because the bonds at the boundary of the tree will have dimension $\leq\chi$, we would require that layer $k+1$ contains at least $6$ nodes. To guarantee this we must have $k \leq \lceil \log{(N)} \rceil - 3$. Therefore we get that for $N$ such that

\begin{equation}
    \lceil \log{(N)} \rceil \geq 4 + \log{(\log{(\chi)})},
\end{equation}

the quantum circuit will contain at least one gate acting on the maximum $4\log{(\chi)}$ qubits. This is validated numerically in Figure \ref{fig:gate_size_heatmap}.

\begin{figure}[htbp]
    \centering
    \begin{subfigure}{0.32\textwidth}
        \centering
        \includegraphics[width=\linewidth]{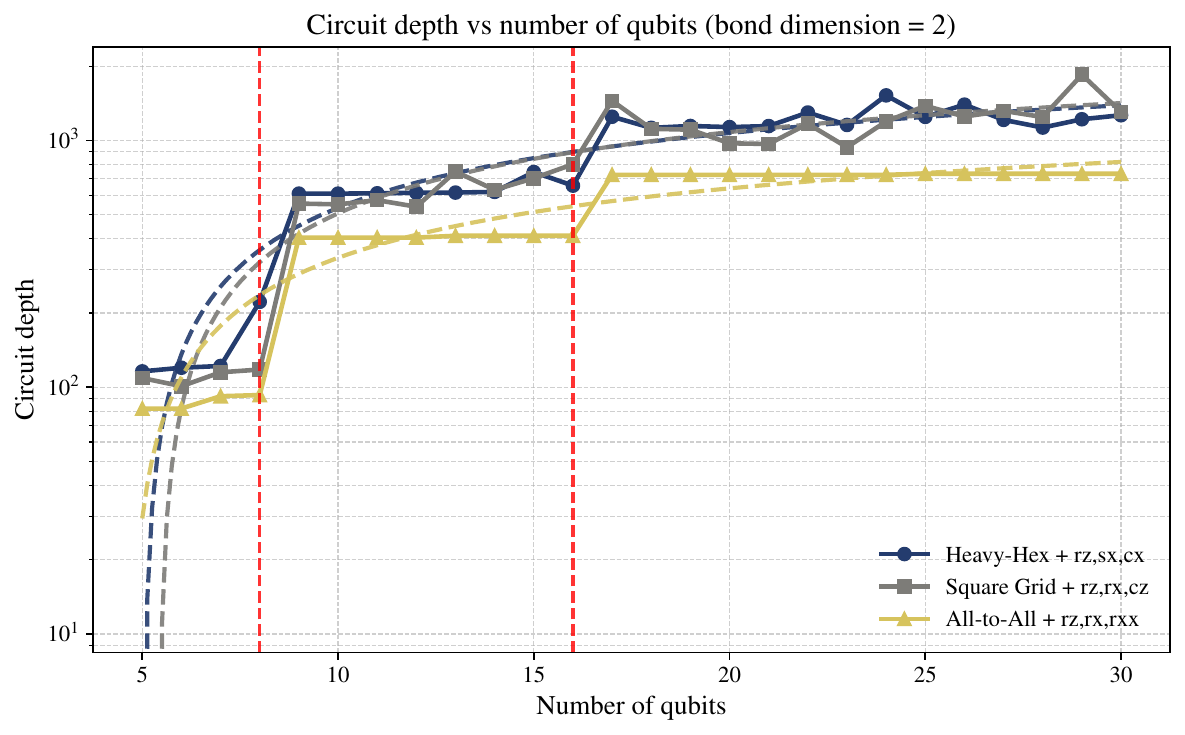}
    \end{subfigure}
    \hfill
    \begin{subfigure}{0.32\textwidth}
        \centering
        \includegraphics[width=\linewidth]{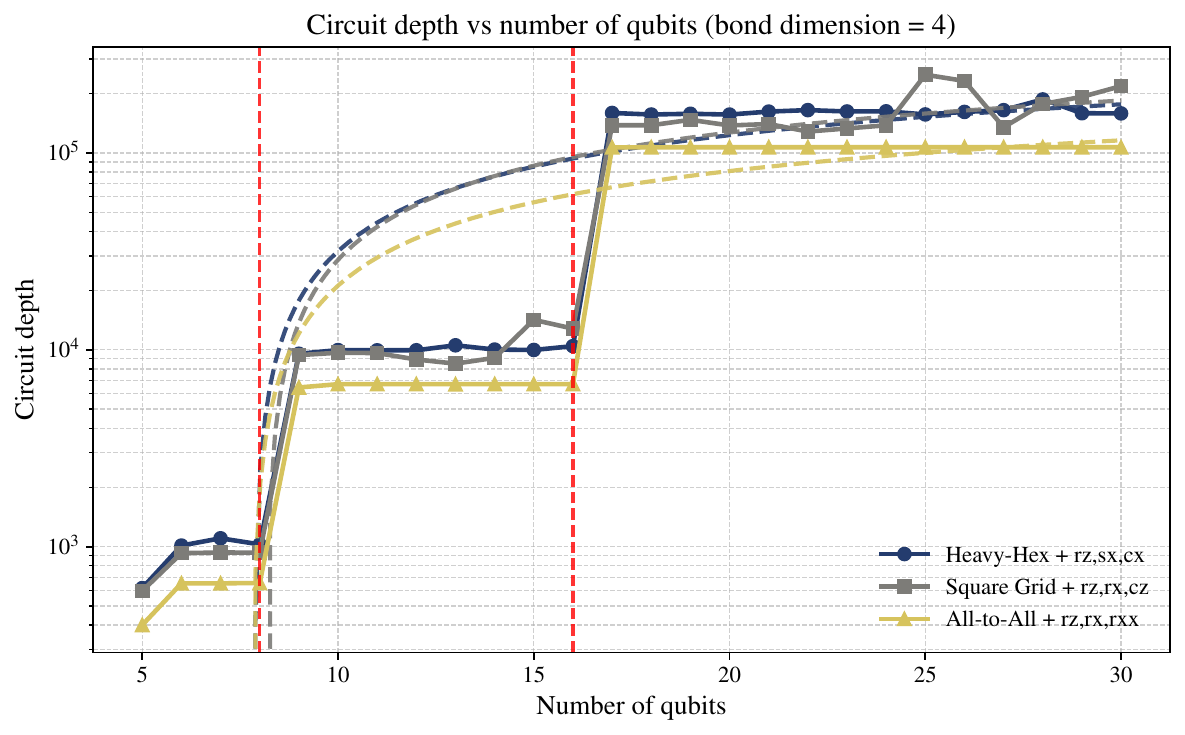}
    \end{subfigure}
    \hfill
    \begin{subfigure}{0.32\textwidth}
        \centering
        \includegraphics[width=\linewidth]{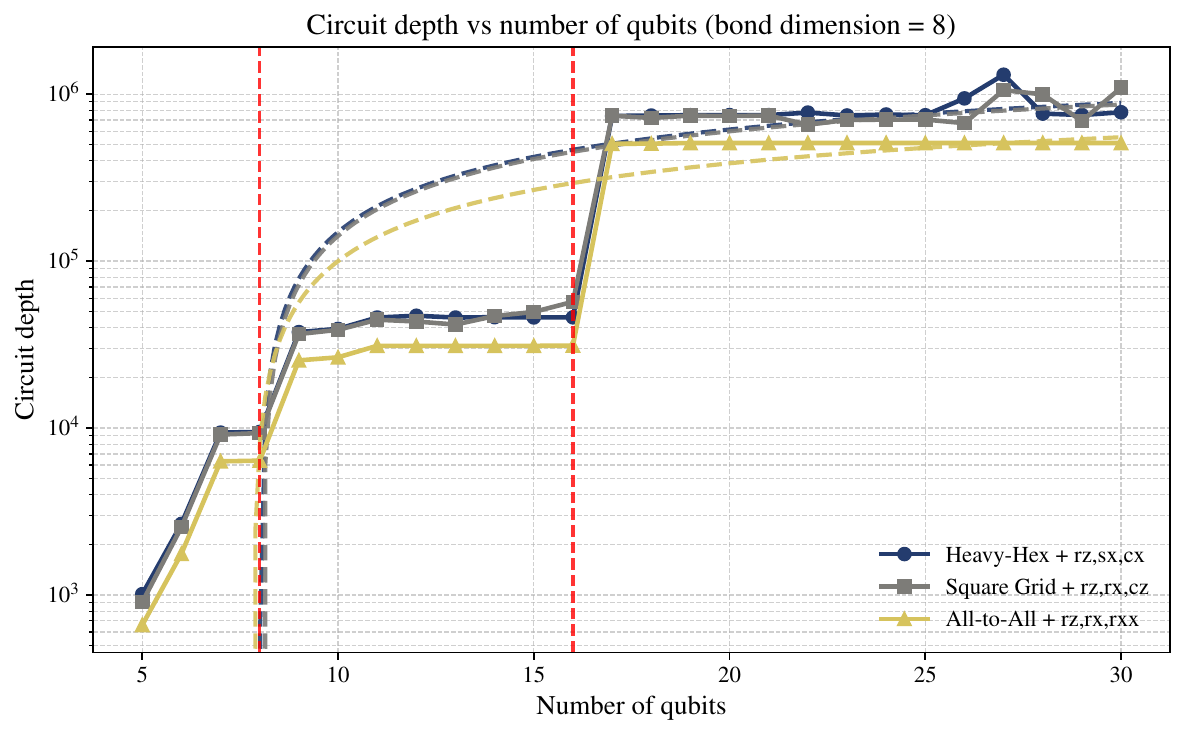}
    \end{subfigure}
    \caption{The circuit depth for exactly preparing randomly generated MPSs with bond dimension 2 (left), 4 (middle), and 8 (right) when transpiled to common quantum architectures. Vertical red dashed lines appear at power-of-2 qubit numbers indicating where an extra layer is required in the TTN and hence the circuit.}
    \label{fig:depth_vs_qubits}
\end{figure}

While the resulting circuits are log-depth asymptotically, there will be prohibitive pre-factors due to the bond dimension growth during the renormalisation procedure. In Figure \ref{fig:depth_vs_qubits} we produce random MPSs with $\chi=2,4,8$ and transpile the resulting quantum circuit to common architectures: all-to-all connectivity and $(R_z, R_x, R_{xx})$ basis gates, square-grid connectivity and $(R_z, R_x, CZ)$ basis gates, and heavy-hex connectivity and $(R_z, S_x, CX)$ basis gates. 

We observe the expected log-depth scaling even when transpiled to restricted topologies, however the transpilation overhead from the multi-qubit gates quickly results in the circuit depths reaching thousands to millions of gates, prohibitive for real quantum hardware. 

\subsection{Approximate Decomposition}
\label{section:approx}

Due to the prohibitive scaling of the exact decomposition, we consider instead an approximate decomopsition achieved by truncating the bond dimension of SVDs. This directly controls the maximum gate size in the final circuits while introducing an approximation error. 

\begin{figure}[htbp]
    \centering
    \begin{subfigure}{0.48\textwidth}
        \centering
        \includegraphics[width=\linewidth]{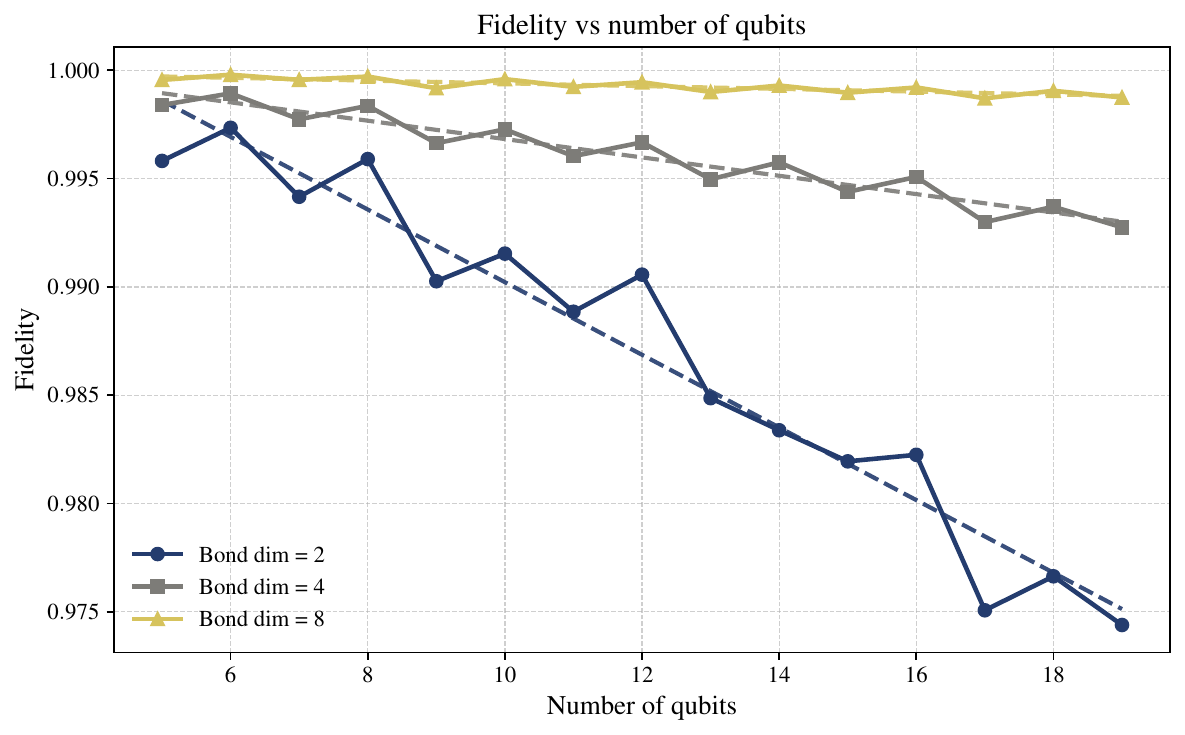}
        \caption{}
    \end{subfigure}
    \hfill
    \begin{subfigure}{0.48\textwidth}
        \centering
        \includegraphics[width=\linewidth]{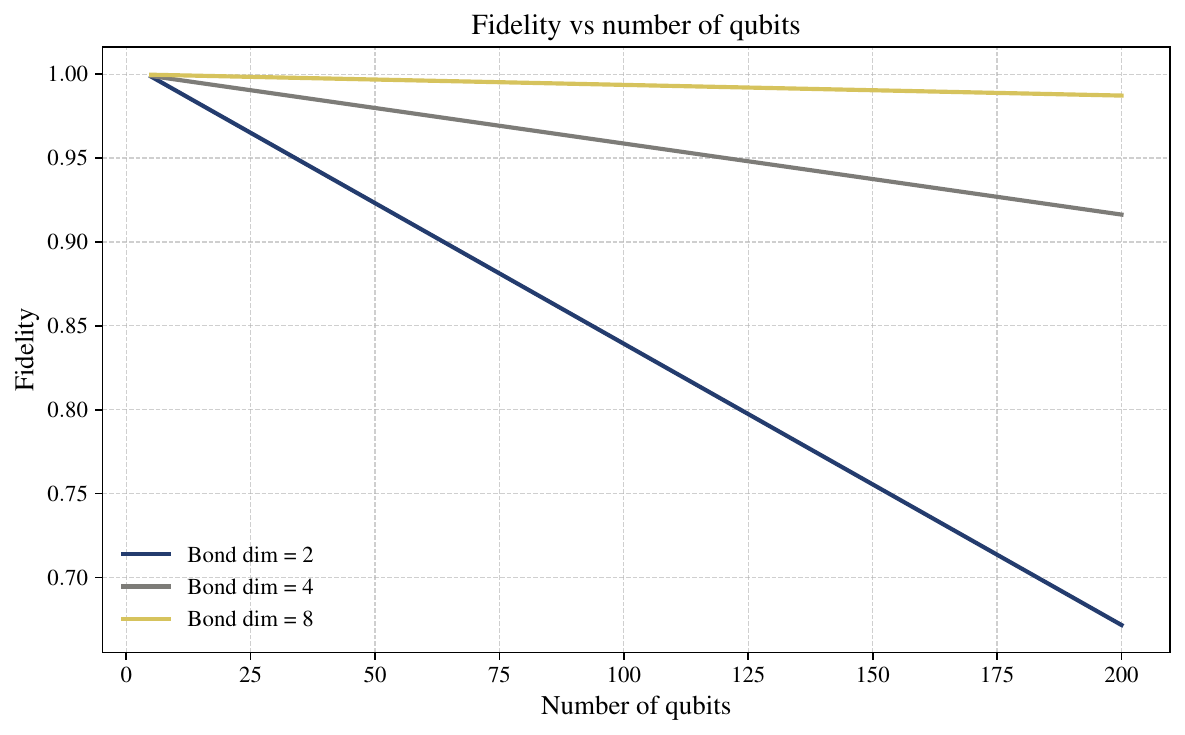}
        \caption{}
    \end{subfigure}
    \caption{The fidelity of the state preparation circuit plotted against number of qubits (left). The linear fitting function is extrapolated to larger numbers of qubits (right).}
    \label{fig:avg_fid_vs_qubits}
\end{figure}

The fidelity / circuit depth trade-off can be tuned for particular instances, however for demonstration purposes we fix the maximum bond dimension to be $2$ so that the final circuits contain only $2$-qubit gates. For various $N$ we randomly generate $30N$ MPSs with bond dimension $\chi=2,4,8$ and produce approximate circuit decompositions. We plot the average fidelity versus $N$ in Figure \ref{fig:avg_fid_vs_qubits}.

We observe that the fidelity decays linearly with $N$. Furthermore, the decay is slow. For up to $20$ qubits the fidelity remains about $0.97$ which is likely to be satisfactory for reference state preparation in algorithms such as QPE and QSCI. Curiously, we note that the approximation obtains significantly higher fidelity for MPSs with larger bond dimension. This is possibly a consequence of using random MPSs, but further investigations should be made with more structured states. 

We extrapolate the linear fit of the fidelity curves (also shown in Figure \ref{fig:avg_fid_vs_qubits}). This shows that even for the worst case $\chi=2$, the fidelity remains high ($>0.8$) for well over $100$ qubits.

\begin{figure}[htbp]
    \centering

    \begin{subfigure}{0.32\textwidth}
        \centering
        \includegraphics[width=\linewidth]{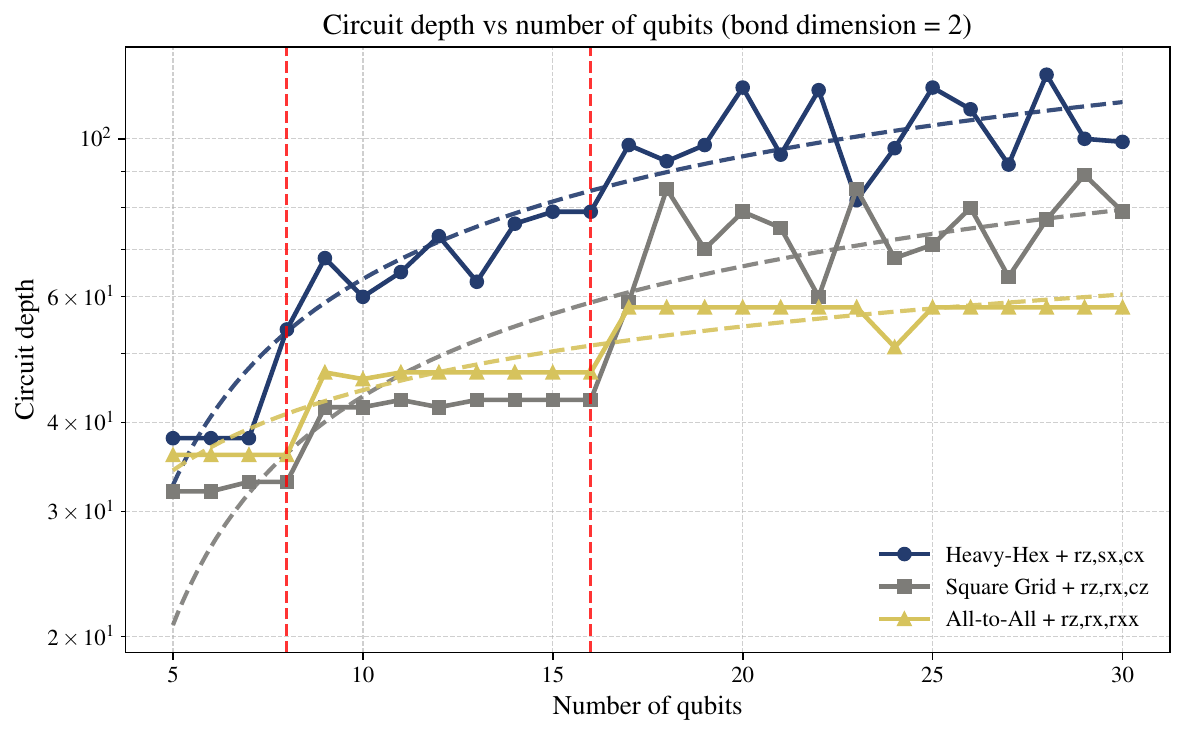}
    \end{subfigure}
    \hfill
    \begin{subfigure}{0.32\textwidth}
        \centering
        \includegraphics[width=\linewidth]{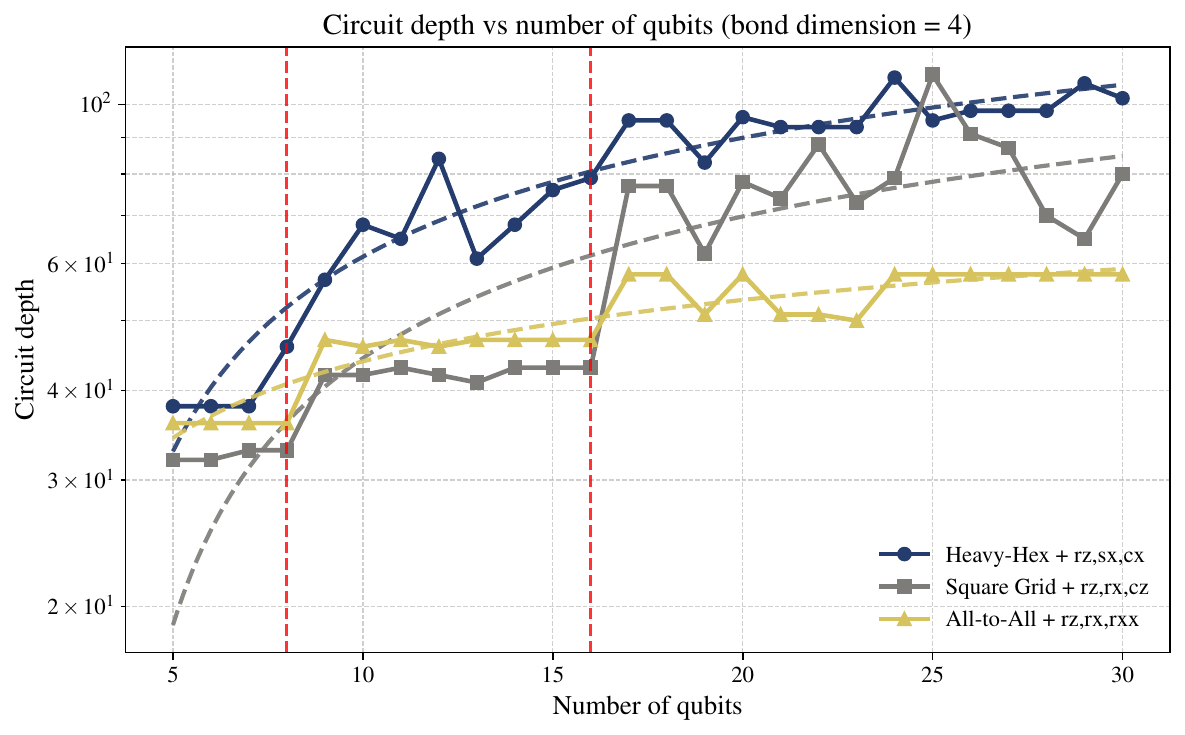}
    \end{subfigure}
    \hfill
    \begin{subfigure}{0.32\textwidth}
        \centering
        \includegraphics[width=\linewidth]{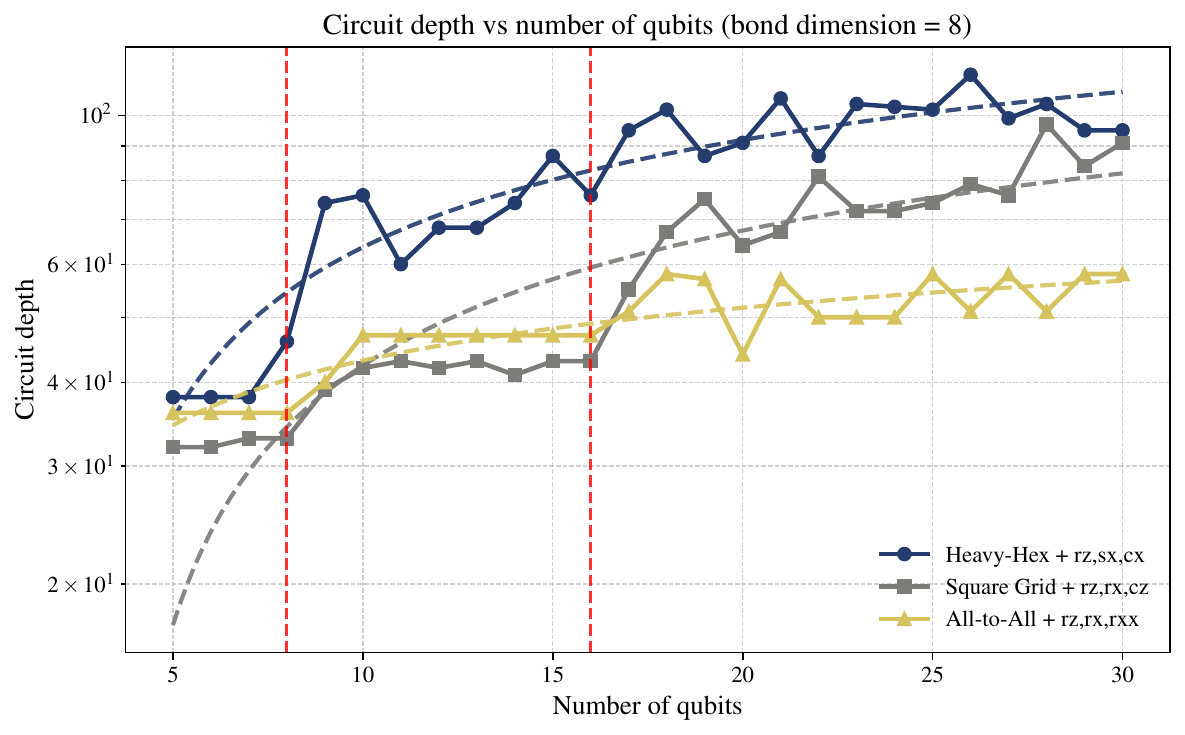}
    \end{subfigure}

    \par\vspace{1em}

    \begin{subfigure}{0.32\textwidth}
        \centering
        \includegraphics[width=\linewidth]{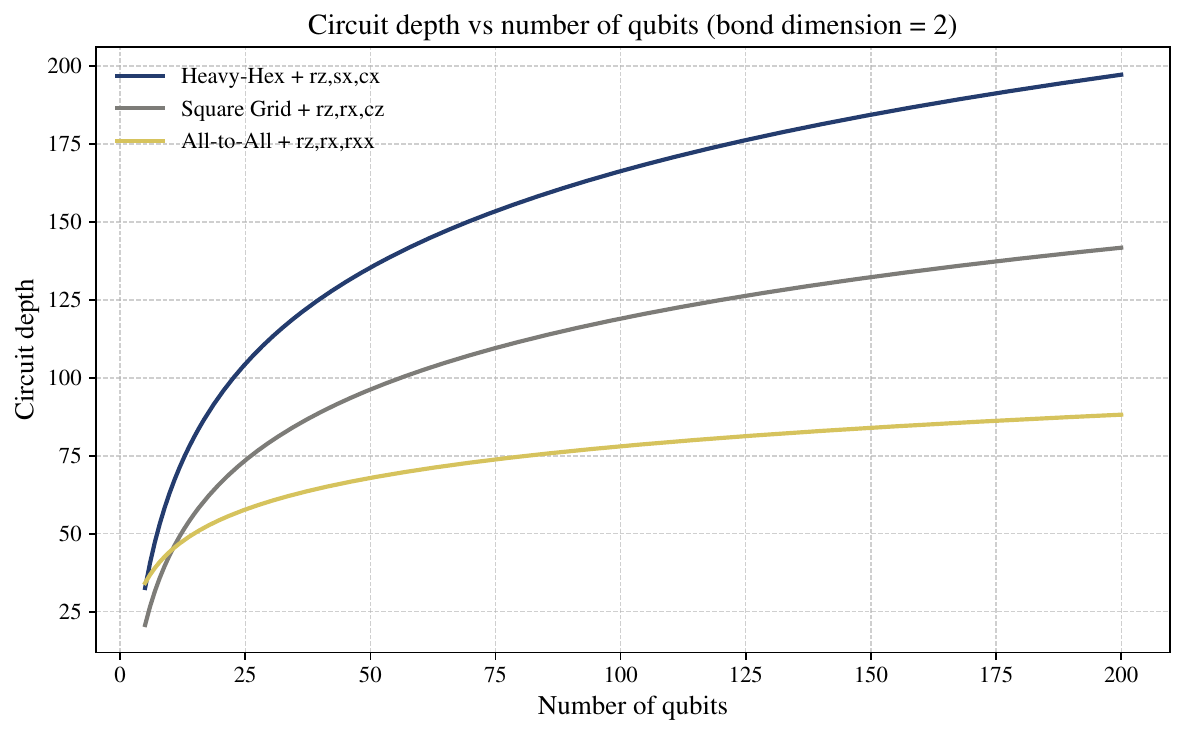}
    \end{subfigure}
    \hfill
    \begin{subfigure}{0.32\textwidth}
        \centering
        \includegraphics[width=\linewidth]{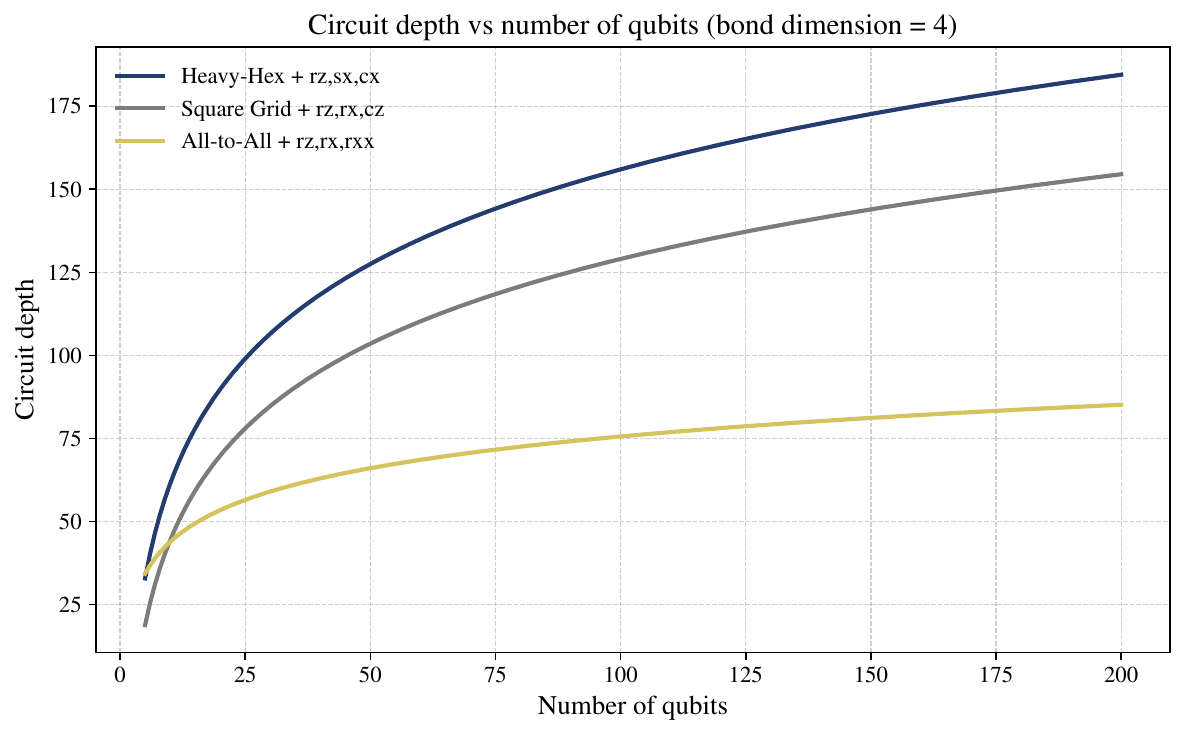}
    \end{subfigure}
    \hfill
    \begin{subfigure}{0.32\textwidth}
        \centering
        \includegraphics[width=\linewidth]{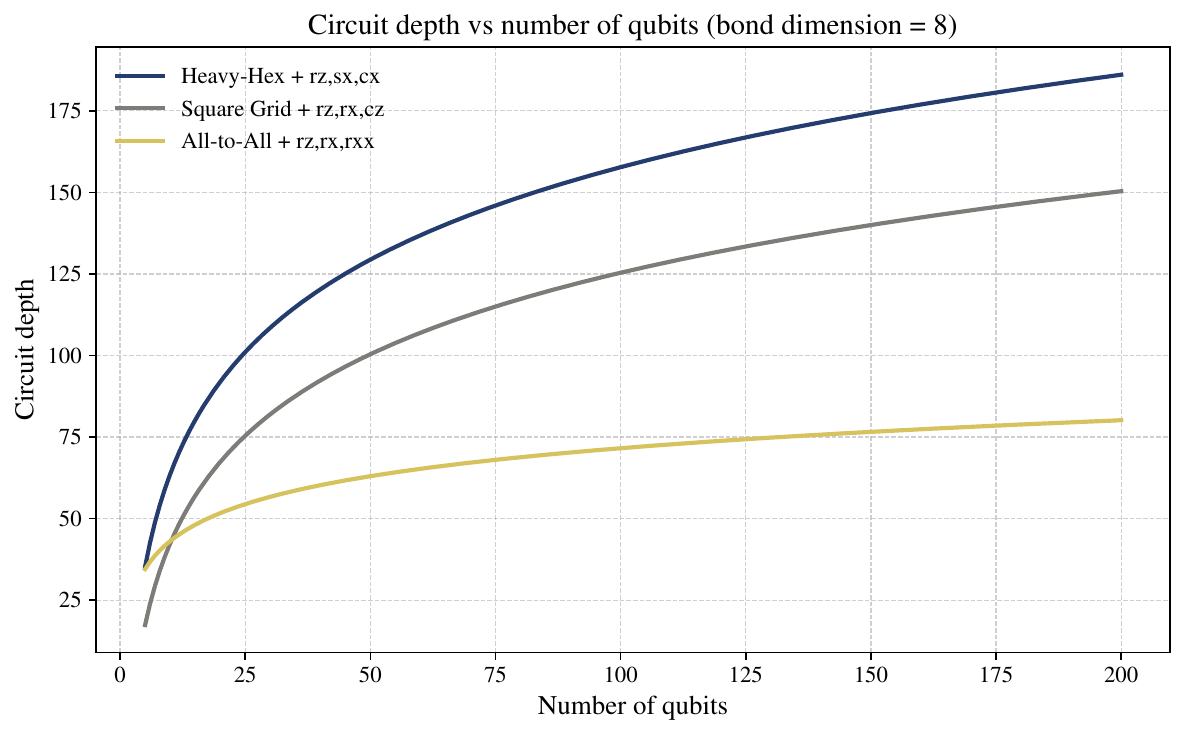}
    \end{subfigure}

    \caption{Top: The circuit depth for approximately preparing randomly generated MPSs with bond dimension 2 (left), 4 (middle), and 8 (right) when transpiled to common quantum architectures. Vertical red dashed lines appear at power-of-2 qubit numbers indicating where an extra layer is required in the TTN and hence the circuit. Bottom: The logarithmic fitting functions extrapolated to larger numbers of qubits.}
    \label{fig:depth_vs_qubits_approx}
\end{figure}

We then repeat the circuit depth study with the approximate decomposition. Transpiling to the same architectures as above, the circuit depths are shown in Figure \ref{fig:depth_vs_qubits_approx} together with the fitting functions extrapolated to larger numbers of qubits.

We observe that even for the most restricted heavy-hex topology we can prepare an MPS with $200$ sites with a circuit depth a little over $200$. For all-to-all connectivity the same MPS can prepared with a circuit depth under $100$. These circuits are therefore much more suited to near-term hardware. 

\section{Verifier Circuits}
\label{section:vc}

We now turn our attention to operators. Consider a unitary operation represented by a matrix product operator (MPO). A natural question is whether we can use a similar decomposition strategy to construct log-depth circuits for MPOs. However, attempting such a decomposition requires either gate sizes scaling with the number of qubits $N$, or local polar decompositions which destroy necessary global structure.

Instead, we can first map the MPO to an MPS through a vectorisation procedure and then perform the state preparation decomposition outlined above. The resulting quantum circuits are examples of \emph{verifier circuits} such as those presented in \cite{mingare2024quantum}. However, here the verifier circuits are log-depth rather than linear-depth, an exponential improvement. 

\subsection{Construction Procedure}

\begin{figure}[htbp]
    \centering
    \includegraphics[width=0.6\textwidth]{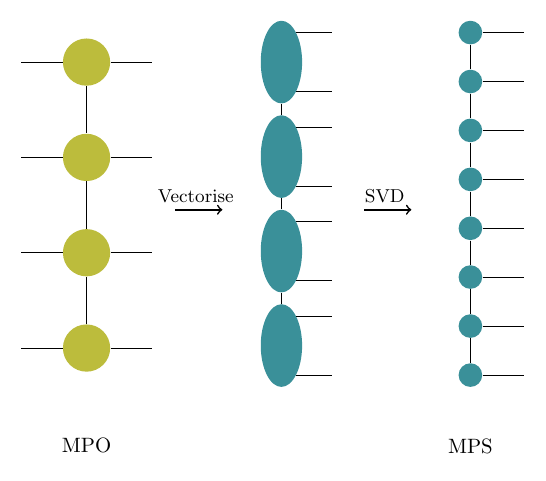}
    \caption{A series of SVDs bring a vecotrised 4-site MPO into the form of an 8-site MPS.}
    \label{fig:mpo_to_mps}
\end{figure}

We consider a unitary operator $U$ represented by an MPO with bond dimension $\chi$. The first step is to transform the MPO into an MPS. The simplest way to do this is to interpret both input and output legs as physical indices and perform a series of SVDs to restore the MPS form as shown in Figure \ref{fig:mpo_to_mps}. The resulting MPS will have $2N$ sites and have bond dimension $2\chi$ in general.

To proceed with the decomposition routine presented in the previous section, we must then normalise the MPS obtained from the MPO. The original norm of $U$ is denoted by $\|U\|$. 

Following the decomposition, we obtain a quantum circuit that implements the operation $V^\dagger_U$ on $2N$ qubits. Inverting the circuit then yields the verifier circuit $V_U$. The interpretation of this circuit is the following. For any input states $\ket{\phi},\ket{\psi}$ we have

\begin{equation}
    \label{eq:verifier-formula}
    \left |\bra{0} V_U \ket{\psi}\ket{\phi}\right |^2 = \frac{1}{\|U\|} \left | \braket{\phi | U | \psi} \right |^2.
\end{equation}

Thus we have constructed a log-depth quantum circuit that can calculate overlaps with respect to the unitary $U$. In particular, if $\ket{\phi} = U\ket{\psi}$ then 

\begin{equation}
    \left |\bra{0} V_U \ket{\psi}\ket{\phi}\right |^2 = \frac{1}{\|U\|} \left |\braket{\psi | U^\dagger U | \psi}\right |^2 = \frac{1}{\|U\|}.
\end{equation}

This explains the name \emph{verifier circuit} in that $V_U$ can be used to text the fidelity of an implementation of a quantum circuit for $U$. 

Alternatively, if $U=I$ then this construction gives a log-depth ancilla-free implementation of the quantum SWAP test. 

\subsection{Fidelity Verification}

We test the properties of our verifier circuit construction using common unitary operators with low-bond dimensoion MPO representations: a multi-control $Z$ gate and the exponential of a single Pauli string both of which have bond dimension $2$. 

\begin{figure}[htbp]
    \centering
    \includegraphics[width=0.5\textwidth]{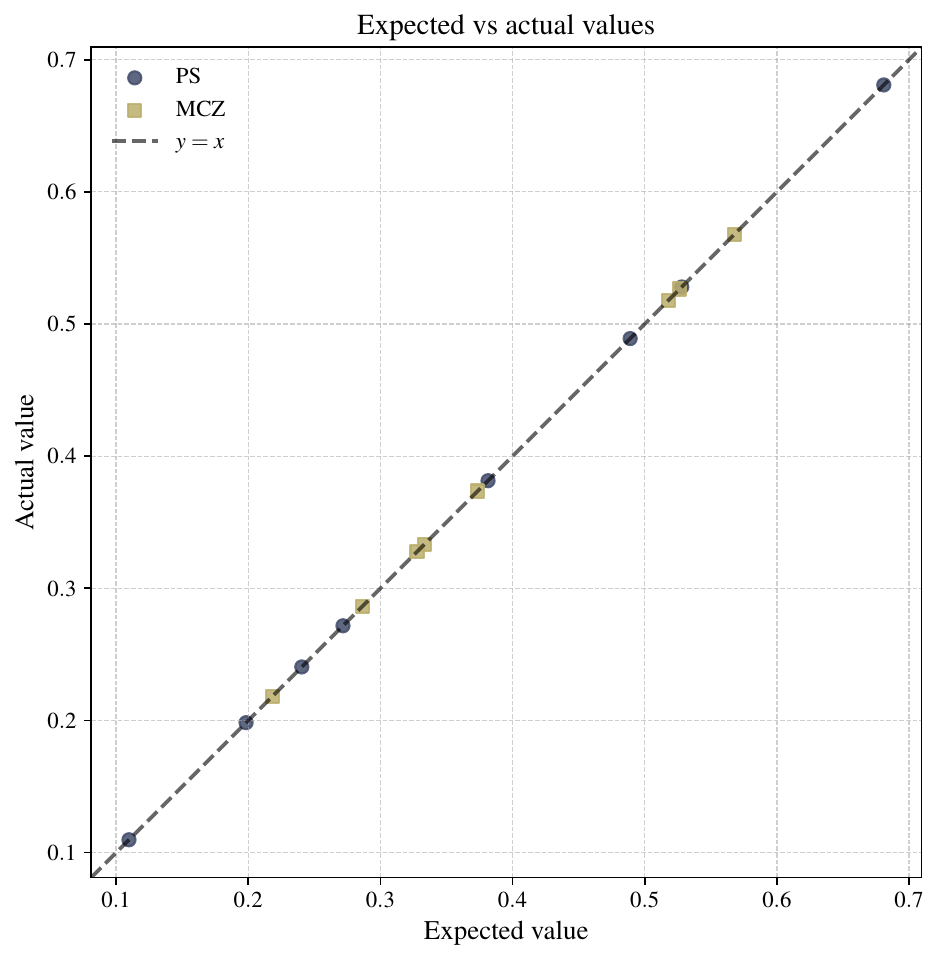}
    \caption{For common unitary operators $U$ we compute the overlap of randomly generated states directly and with the verifier circuit construction. The exact results free of shot noise are compared in this plot and shown to lie on the line $y=x$.}
    \label{fig:vc_fid}
\end{figure}

For randomly sampled input states, we compute directly the overlaps $\braket{\phi | U | \psi}$ as well as the amplitude $\braket{0|V_U|\psi}\ket{\phi}$ to validate Equation \ref{eq:verifier-formula}. The results are shown in Figure \ref{fig:vc_fid} and show the expected relationship.

\begin{figure}[htbp]
    \centering
    \includegraphics[width=0.5\textwidth]{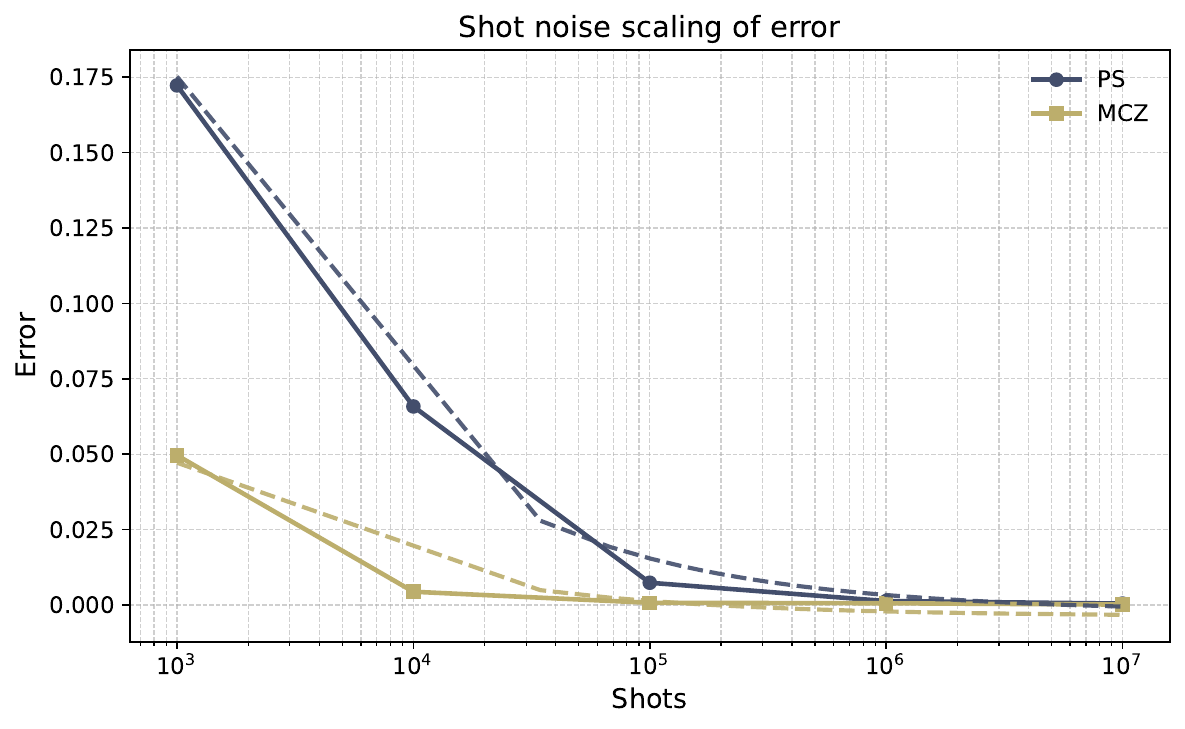}
    \caption{The error of the verifier circuit construction plotted as a function of number of shots.}
    \label{fig:vc_shots}
\end{figure}

In practice, the amplitude $\left |\braket{0|V_U|\psi}\ket{\phi}\right |^2$ will be measured with a finite number of shots $S$, leading to the usual error scaling $\epsilon \propto 1/\sqrt{S}$ as demonstrated in Figure \ref{fig:vc_shots}. 

\subsection{Noise Sensitivity}

The verifier circuit construction may be useful when a problem reduces to calculating an overlap magnitude, $\left | \braket{\phi|U|\psi} \right |$, and when the unitary $U$ has a relatively efficient description as an MPO. One specific application is to circuit-level device calibration. That is, the verifier circuit $V_U$ provides a metric for the fidelity of an implementation of the circuit $U$ on quantum hardware. 

\begin{figure}[htbp]
    \centering
    \includegraphics[width=0.5\textwidth]{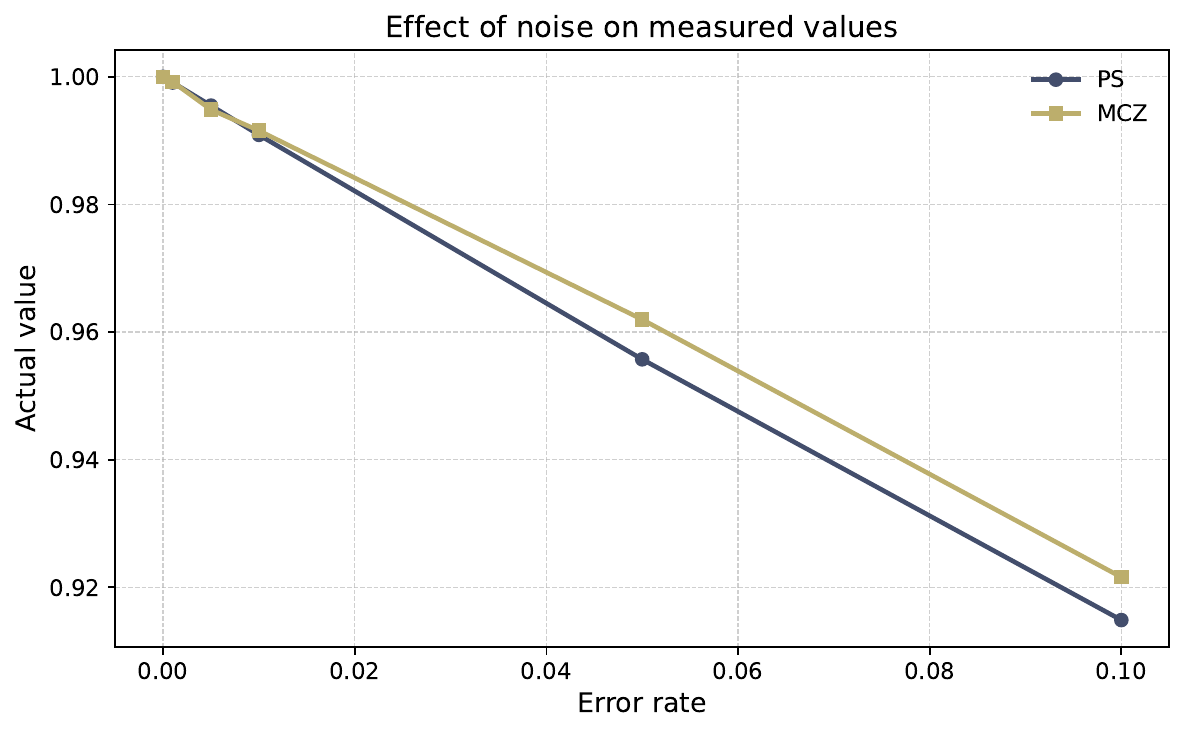}
    \caption{The verifier circuit provides a fidelity metric for circuit implementations that is shown to decrease monotonically as noise increases.}
    \label{fig:vc_noise}
\end{figure}

Suppose we have a quantum circuit $U$ with a low-bond dimension MPO representation. Its implementation on noisy hardware is denoted $\tilde{U}$. Prepare two copoies of an arbitrary input state, $\ket{\psi}$, and evolve one by $\tilde{U}$. Then use these two states as input to the verifier circuit $V_U$ so that the amplitude of the all zero state gives,

\begin{equation}
    \left |\braket{0 | V_U | \psi} \tilde{U} \ket{\psi}\right |^2 = \frac{1}{\|U\|} \left |\braket{\psi | \tilde{U}^\dagger U | \psi}\right |^2 \leq \frac{1}{\|U\|},
\end{equation}

with equality if and only if $\tilde{U} = U$. To be a useful device calibration metric we require that the measured amplitude decreases monotonically as the device noise increases. We can test this by defining a noise parameter $0 \leq\delta\leq 1$ and constructing the initial state

\begin{equation}
    \ket{\psi} \otimes \tilde{U} \ket{\psi} = \ket{\psi} \otimes (\sqrt{1-\delta} U \ket{\psi} + \sqrt{\delta} \ket{\eta}),
\end{equation}

where $\ket{\eta}$ is a random noise perturbation to the expected state. We then plot $\|U\| \left |\bra{0} V_U \ket{\psi}\ket{\phi}\right |^2$ against $\delta$ in Figure \ref{fig:vc_noise}.

We observe a linear decrease in fidelity as the noise increases implying that the verifier circuit is indeed a useful construction for circuit-level device calibration. 

\section{Conclusion}

We have presented a conceptually simple and practical method for decomposing MPSs into log-depth state preparation circuits with a controllable fidelity / circuit depth trade-off. This provides a route to MPS warm-started quantum algorithms such as using the DMRG solution as a reference state for quantum phase estimation or quantum-selected configuration interaction. 

We extended this method to MPOs resulting in the construction of log-depth and ancilla-free circuits to compute overlaps of the form $\left | \braket{\phi|U|\psi} \right |^2$ given arbitrary states $\ket{\phi},\ket{\psi}$. We suggest a use for this decomposition for circuit-level device calibration, however other applications based on this construction may exist and are left to future work.

\section{Conflicts of Interest}

The authors declare no conflicts of interest.

\section{Data Availability}

Our method is implemented in the open-source repository TN4QA (available through PyPI or directly here: https://github.com/UCL-CCS/TN4QA). The notebook used to produce the results presented in this work is avaiable upon request to the corresponding author. 

\section{Acknowledgments}

We acknowledge support from the Engineering and Physical Sciences Research Council (EPSRC, grant number EP/T517793/1). PVC is grateful for funding from the European Commission for VECMA (800925) and EPSRC for SEAVEA (EP/W007711/1). 

\bibliography{bib}

\end{document}